# A compact resonant Π-shaped photoacoustic cell with low window background for gas sensing

A. L. Ulasevich[1], A. V. Gorelik[1], A. A. Kouzmouk[1], V. S. Starovoitov[1,2]

[1]*B.I.Stepanov Institute of Physics, NASB, Nezavisimosti ave. 68, 220072 Minsk, Belarus*
[2] *Department of Physics, Humboldt University, Berlin, Newtonstrasse 15, D-12489 Berlin, Germany*
e-mail:  starovoitov@physik.hu-berlin.de

A resonant photoacoustic cell capable of detecting the traces of gases at an amplitude-modulation regime is represented. The cell is designed so as to minimize the window background for the cell operation at a selected acoustic resonance. A compact prototype cell (the volume of acoustic cavity of ~ 0.2 cm$^3$, total cell weight of 3.5 g) adapted to the narrow diffraction-limited beam of near-infrared laser is produced and examined experimentally. The noise-associated measurement error and laser-initiated signals are studied as functions of modulation frequency. The background signal and useful response to light absorption by the gas are analyzed in measurements of absorption for ammonia traces in nitrogen flow with the help of a pigtailed DFB laser diode operated near a wavelength of 1.53 μm. The performance of absorption detection and gas-leak sensing for the prototype operated at the second longitudinal acoustic resonance (the resonance frequency of ~ 4.38 kHz, $Q$-factor of ~ 13.9) is estimated. The noise-equivalent absorption normalized to laser-beam power and detection bandwidth is ~ 1.44×10$^{-9}$ cm$^{-1}$ W Hz$^{-1/2}$. The amplitude of the window-background signal is equivalent to an absorption coefficient of ~ 2.82×10$^{-7}$ cm$^{-1}$.

## 1 Introduction

Laser photoacoustic spectroscopy is an efficient technique to be applied to local non-contact analysis of trace amounts of various chemical compounds in gas media [1-4]. The principle of the technique is based on measuring the amplitude and phase for an acoustic pressure oscillation (a so-called photoacoustic response or signal) arising due to absorption of a modulated laser beam by molecules of gas inside a photoacoustic cell. The photoacoustic gas detection is realized with an enhanced sensitivity if the modulation frequency coincides with an acoustic resonance of the internal cell cavity. The properly designed resonant cell is specified by a low level of window-background signal (a photoacoustic response arisen in the cell due to absorption and reflection of light beam by the cell windows) for the selected acoustic resonance. The resonant photoacoustic technique based on infrared lasers is distinguished by a high sensitivity (the minimal detectable absorption of 10$^{-10}$ cm$^{-1}$ can be attained at a time resolution of a few seconds) and capability to recognize reliably a large number of chemical compounds [5-8]. The technique has found successful exploitation for a large number of practical applications [9-13].

A promising line of development for the photoacoustic technique is associated with creating a miniature resonant cell [14,15]. According to theoretical estimations [14, 16, 17], the amplitude of photoacoustic response can be increased with reducing the cell sizes. The miniature high-sensitivity photoacoustic cell is a valuable facility in order to analyze chemical compounds to be emitted by individual small-sized objects at an extremely low emission rate. The compact photoacoustic gas sensor can be successfully exploited as a gas-leak detector. In accordance with our crude estimation, the laser photoacoustic sensor equipped with a resonant cell, the cavity volume of which is a few cubic millimeters, can demonstrate a substantially higher sensitivity of gas-leak detection in comparison to the best-performance commercial gas-leak-detection systems [18-22]. The photoacoustic leak detector can be applied to *in situ* localization of leak for a large number of gases to be emitted in atmospheric air. The most appropriate light sources for the compact photoacoustic gas sensor are miniature single-mode semiconductor lasers operated in the near- or mid-infrared wavelength region. They are laser diodes (the oscillation wavelength region of 0.4 - 4 μm) or quantum cascade lasers (4 - 20 μm) [23-25].

In practice, there are a few efficient approaches to miniaturize the resonant photoacoustic cell. The quartz-enhanced photoacoustic spectroscopy (QEPAS) is a well-developed way to the miniature gas sensor [26]. Instead of a gas-filled resonant acoustic cavity, the sound energy is accumulated in a high-Q quartz crystal frequency standard. Usually, the standard is a quartz tuning fork with an acoustic resonant frequency of 32 kHz in air. A theoretical model (that enables the detected piezoelectric signal to be expressed in terms of optical, mechanical, and electrical parameters of the QEPAS sensor) has been developed [27]. A great success is achieved in QEPAS-based gas detection with the help of different infrared laser systems (including laser diode, quantum cascade lasers and optical parametric oscillators) [28-32]. The volume of acoustic cavity for the developed cells reaches down to a few cubic centimeters. The QEPAS-based experiments demonstrate that the high-sensitivity photoacoustic detection can be realized at ultrasonic modulation frequencies if the rates of intra- and intermolecular collisional vibration-vibration (VV) or vibration-translation (VT) energy redistribution for the detected species are high compared to the modulation frequency.

The window background has a detrimental slightly-wavelength-dependent effect on the gas-detection performance of gas sensor. The design of QEPAS sensor does not imply any special protection measures against this effect [33]. In order to minimize the background effect, the QEPAS sensor has to operate at a wavelength-modulation regime (a modulation regime insensitive to the wavelength-independent signals). The sensors operated at this regime are applicable for detection of gases (usually, they are molecules of a few atoms) with narrow spectral absorption lines. But, these sensors cannot be applied efficiently for detection of compounds (complex polyatomic molecules) with broad spectral absorption bands (such bands are very similar to the window background). The application potential of the QEPAS approach is restricted also due to non-flexible and complicated design of the photoacoustic cell: the laser beam should be passed without touching sequentially through long and narrow tubes (they are attached to the fork) and a narrow gap (usually, this gap is 300 μm) between the fork prongs.

Another approach to the cell miniaturization implies a traditional design of photoacoustic cell: through a small hole in the cell shell, a standard broadband acoustic sensor (for instance, a condenser microphone or an ultrasonic transducer) registers response to the light beam modulated with the frequency of an acoustic resonance of cell cavity. The cell design can be optimized in order to provide the best gas-detection performance for the selected acoustic resonance. The optimization allows one to minimize the parasite acoustic signals (the window background, the noise to be initiated by external acoustic disturbances), which can play an extremely great negative role for the small-sized cells [34, 35]. For the properly optimized cell, the window background is absent at any modulation regime. Therefore, such a cell is applicable to detection of any infrared-active chemical compound whatever the spectral features of the compound. This cell operated at an amplitude-modulation regime (a modulation regime, which is sensitive to the wavelength-independent signals) can be used successfully in order to detect simple compounds with narrow spectral absorption lines or to sense the traces of complex compounds with broad absorption bands. The performance of traditionally-designed cell is not critical to variation in pressure and temperature of the gas to be analyzed. The cell does not need a special equipment to control the gas flow through the cell. The cell can be easily adapted to any laser-beam diameter and modulation frequency.

The traditionally-designed photoacoustic cell can be miniaturized down to microelectromechanical-system (MEMS) dimensions by a trivial size-scaling procedure applied to a well-developed macro-scale cell [36, 37]. The potential of MEMS-scaled photoacoustic cells both for gas detection at a parts-per-billion-concentration level and for multicomponent gas analysis was demonstrated in experiments [38, 39]. Recently, the size-scaling procedure has been used in order to produce a miniature prototype of banana-shaped cell [40] operated at a longitudinal acoustic mode [41, 42]. In spite of low gas-sensing performance (the noise-limited minimal detectable absorption normalized to light-beam power and detection bandwidth is ~$8.07 \times 10^{-8}$ cm$^{-1}$ W Hz$^{-1/2}$), the produced prototype cell shows a good capability of gas-leak detection [41]. The noise-limited minimal ammonia-leak rate of $\sim 6.0 \times 10^{-8}$ cm$^3$/s, which can be attained with the help of the prototype at a time resolution of 1 s and a typical light-beam power of 10 mW, is considerably better compared to the relevant rate to be demonstrated by commercial hand-held halogen/hydrogen/helium sniffer leak detectors [20-22]. A deciding factor for such an increase in leak-detection sensitivity is the reduced cell sizes (the volume of acoustic cavity for the prototype is ~5 mm$^3$). A considerably high level of background signal (the absorption equivalent of window background is $\sim 2.5 \times 10^{-5}$ cm$^{-1}$ at a resonance-peak frequency) demonstrates a non-ideal optimization for the cell design.

The most accurate way of cell optimization can be performed with the help of a numeric simulation, which estimates correctly the spatial acoustic signals to be generated inside the cell. The standard finite-element method based on a direct numerical solution of the acoustic Helmholtz equation is a reliable well-proven technique to be used for optimizing the design of photoacoustic cell [43-45]. A finite-element tool to be applied usually for such purposes is the COMSOL Multiphysics software [46].

In the work, we present a resonant photoacoustic cell. Like the banana-shaped cells [40, 41], our cell is intended for operation with a collimated linear-polarized light beam. The cell is designed and produced in accordance with suggestions disclosed in [41] and directed, in general, to enhance the performance of miniaturized resonant cell. With respect to the banana-like shape, the shape of acoustic cavity of the presented cell is somewhat modified. The design modification facilitates the process of cell production and reduces the effect of imperfections associated with the cell production. The parameters, which specify the cavity design, are chosen so as to minimize the imperfection of cavity design and to provide the optimal cell operation at a selected longitudinal acoustic resonance. The parameters are chosen with the help of numeric simulation using a standard finite-element method.

The presented design of cell cavity is implemented in a compact prototype cell made of a light-weight and low-cost plastics. Like the cell described in [41], the produced prototype cell is adapted to the narrow low-divergence light beam to be generated by a near-infrared semiconductor laser. In comparison to the cell [41], our prototype is designed for a higher absorption-detection sensitivity. The sensitivity enhancement is accompanied by an increase in the cell sizes. The performance of the produced prototype is evaluated in an experimental examination. In the examination we study the frequency spectrum of measurement error associated with acoustic-sensor noise. Then we analyze the amplitude-frequency dependence for photoacoustic signals (they are the window background and useful response to light absorption by the gas) to be initiated by the laser beam inside the prototype cell. The parameters, which specify the gas-detection performance of prototype for the selected acoustic resonance, are estimated and compared with the relevant parameters of banana-shaped cells [40, 41]. In this estimation we evaluate the capability of our cell in detection of absorption in gas. This function is very important for measuring the concentration of chemical compounds in gas media. Also, we estimate the performance for the cell to sense gas leaks emitted locally by individual objects.

# 2 Photoacoustic cell

## 2.1 Basic design of cell cavity

Here we describe a design of acoustic cavity for the resonant photoacoustic cell. The cell is intended for detection of gases with the help of a collimated and linearly polarized light beam. As for the banana-shaped cells [40, 41], the acoustic cavity of our cell consists of three adjacent cylindrical parts (one central and two lateral cylinders). Our cell is distinct from the cells [40, 41] by the cavity shape. The arrangement of the cavity parts resembles the capital Greek letter "Π". Such a simple "Π"-like cavity shape is chosen for the purpose of facilitating the process of cell production. The cross-section diameters for the central and lateral cylinders are identical and equal to $D$. The symmetry axes of all three cylindrical cavities are located on the plane $OO'P$ formed by the optical axis $OO'$ and electric polarization vector $\boldsymbol{P}$ for the light beam. Fig. 1 shows a section of the acoustic cavity of cell by the plane $OO'P$.

The light beam passes through the central cylindrical part along the optical axis $OO'$, which coincides with the axis of cylindrical symmetry for this cavity part. The points $O$ and $O'$ are points of intersection of the axis $OO'$ and the internal surfaces of optical front- and back-end windows of the cell. In order to minimize the parasite reflection of light beam on the window surfaces the front- and back-end windows are mounted at the Brewster angle $\Theta_B$ relative to the axis $OO'$ ($\Theta_B$ is the angle between $OO'$ and the normal to the window surface). The distance $L_C$ between the points $O$ and $O'$ is accepted as a length of central cylindrical part.

The lateral cylindrical parts are identical in the sizes. The parts are located symmetrically relative to the central part. The symmetry axes of these parts go through the point $O$ or $O'$. Each of these axes is perpendicular to the axis $OO'$. For each lateral cylindrical parts, the distance $L_{Lat}$ between the point $O$ (or $O'$) and internal surface of the end along the cylindrical-symmetry axis is accepted as a cylinder length.

The design of cavity for our Π-shaped cell is optimized in order to enhance the cell performance for a selected longitudinal acoustic resonance. The optimization minimizes the window-background signal and, simultaneously, provides a high level for the useful response. The optimization is performed with the help of a numeric simulation based on a standard finite-element method (eigenfrequency analysis of pressure acoustics in "Acoustics" module of COMSOL Multiphysics 3.5 software [46]). The simulation allows one to determine the spatial distribution of acoustic pressure wave inside the cavity at the resonance. At the optimization we adapt the cell cavity to a required optimal acoustic pressure distribution by tuning a parameter, which specifies the cavity shape. The length $L_{Lat}$ is such a tunable parameter. At the optimization the diameter $D$, length $L_C$ and angle $\Theta_B$ take on given fixed values.

Like the banana-shaped cells [40, 41], our cell is optimized for operation at the second longitudinal acoustic mode (denoted as a $v_2$ mode) of cell cavity. As for the cells [40, 41], the cavity shape of our cell is optimal (the window background is minimal at a high amplitude of useful response) if the length of lateral cylindrical parts $L_{Lat}$ is close to half the length $L_C$. At such a length proportion the nodes of $v_2$ mode are located in the vicinity of the front- and back-end cell windows. We accept that the parameter $L_{Lat}$ takes on an accurate optimal value if the nodes of spatial pressure distribution simulated for $v_2$ mode at this parameter value are located in the points $O$ and $O'$. The inlet and outlet gas holes are sited near the node surfaces of $v_2$ mode. Such a position of the gas holes is intended to reduce the negative influence of holes on the acoustic $Q$-factor and, simultaneously, to isolate the measurement from external acoustic noise for the $v_2$ mode. The inlet/outlet gas holes are connected by ducts to nipples. The ducts are directed up perpendicularly to the plane $OO'P$. The nipples are adapted to flexible gas tubing. The photoacoustic response is registered by an acoustic sensor $M$ located near the midpoint of the central cavity part.

## 2.2 Cell prototype

The Π-shaped design of cell cavity is implemented in a produced compact prototype of photoacoustic cell. The shell of the cell prototype is made of PIC plastics (a polyvinylchloride plastics). The PIC plastics is a light-weight and hard material, which is chemically resistant to a large number of gases and vapours. The prototype is adapted to the narrow diffraction-limited light beam to be generated by a near-infrared semiconductor laser. The length of central cylindrical part $L_C$ is 39 mm. The cross-section diameter $D$ for the central and lateral cylinders is 1.8 mm. The optical windows are made of $CaF_2$. The angle $\Theta_B$ is 55 degree. Such an angle value is equal approximately to the Brewster angle for $CaF_2$ over a broad near-infrared wavelength range [47]. The length $L_{Lat}$ is 19.7 mm. This quantity answers to an optimal acoustic pressure distribution (the pressure nodes are located in the points $O$ and $O'$) simulated for $v_2$ mode at the given values of $D$, $L_C$ and $\Theta_B$. According to the simulated data, inside the so optimized cavity, the nodes surfaces of $v_2$ mode are close to planes, which are intersections of the central and lateral cavity parts.

The acoustic sensor M mounted in the cell shell is a miniature condenser Knowles FG-3629 microphone (sensitivity ~ -53 dB, noise calculated to SPL ~ 28 dB at 1 kHz). The microphone is connected to the acoustic cell cavity by a duct (the duct diameter and length are, correspondingly, 0.7 and 1 mm). The cross-section diameter for the inlet and outlet holes in the cell shell is 0.3 mm. The inlet/outlet nipples are standard SMC one-touch fittings adapted to a tubing with the internal cross-section diameter of 1.2 mm. The volume $V$ of acoustic cavity for the prototype is approximately 194 mm$^3$. The total cell weight (including the cell shell, windows, nipples, microphone and electric wiring) is 3.5 g. According to our estimation, the eigen-frequency of $v_2$ mode for the cavity prototype is approximately 4.43 kHz. The photo of the prototype cell is represented in Fig. 2.

# 3 Experimental details

## 3.1 Procedure of noise-associated measurement-error estimation

In the experiment, we estimate a measurement error for the detected photoacoustic response due to noise to be generated by the microphone inside the prototype cell in the absence of light beam. We apply an error-estimation technique, which is similar to a procedure described in [35, 41].

The prototype cell is inserted into a capsule. The capsule provides a reliable sound isolation for the cell cavity from the noise produced in our laboratory room. The inlet and outlet gas ducts of the cell are closed. The cell cavity is filled with conditioned laboratory air. A regime of strict silence is kept in the laboratory room. The measurement error $\sigma_f$ to be obtained at such an acoustic-isolation regime corresponds to the minimal error level attainable by the microphone signal.

The voltage signal $S_t$ from the cell microphone is amplified by a frequency-selective low-noise amplifier (selective nanovoltmeter type 273, Unipan) and stored by a HS3 digital oscilloscope connected to a personal computer. The microphone signal is stored by the oscilloscope as a time-sample signal realization over a fixed time interval $\tau_l \approx 1.31$ s. The photoacoustic response is determined as a Fourier transform $S_f$ of the microphone signal $S_t$. The quantity $S_f$ is a complex-valued function of frequency $f$ calculated with the help of a fast-Fourier-transform procedure performed for an individual time-sample signal realization stored in the oscilloscope. The time of signal averaging $\tau_{avr}$ for the quantity $S_f$ is equal to the time of signal realization $\tau_l$. The value of error for each frequency component of quantity $S_f$ is determined as a bandwidth-normalized standard root-mean-square deviation of $S_f$:

$$\sigma_f = \tau_{avr}^{1/2} \, (<S_f S_f^*> - <S_f><S_f^*>)^{1/2}. \tag{1}$$

The symbol $<...>$ means the averaging over an ensemble of the time-sample signal realizations. The number of the signal-sample realizations used for the ensemble-averaging procedure is not less than 1000. We evaluate the noise-associated measurement error for each frequency component of photoacoustic response and analyze the frequency spectrum of deviation $\sigma_f$.

## 3.2 Experimental setup

The laser-initiated photoacoustic signals arisen inside the presented photoacoustic prototype cell are analyzed by an experimental way. The used experimental setup is shown in Fig. 3. In the experiment, the prototype is examined with the help of a collimated near-infrared light beam. The beam is generated by a current-modulated laser source. The source is a fiber pigtailed single-mode distributed-feedback laser diode (D2547PG57, Agere Laser 2000) operated near a wavelength of 1.53 μm. The current for the laser diode is supplied by a Thorlabs current controller LDC 202C. The laser diode operates at an amplitude current-modulation regime. The current modulation is performed with the help of a TTL-like square-wave reference signal directed from a Handyscope HS3 digital oscilloscope to the current controller. The frequency $f_m$ of reference-signal switching is accepted as a frequency of beam modulation. A Thorlabs TED 350 temperature controller is used in order to maintain the laser diode at a fixed temperature.

The laser beam is output from a PM-fiber and a Thorlabs CFC-2X-C collimator and directed through a half-wave quartz plate and the prototype cell to a power meter (Ophir PM 3A). The half-wave plate is optimized for operation near a wavelength of 1.535 μm. The plate is used for fine orientation tuning of the beam polarization vector **P** along the plane formed by the symmetry axes of all three cylindrical cavities of prototype cell. The efficient cross-section diameter of the collimated laser beam, which passes the central cavity part of the prototype, is ~ 0.38 mm. The prototype cell is thoroughly adjusted in such a way as to provide the best transmission of the beam through the prototype. The laser-beam power $P_{on}$, which is measured by the power meter immediately after the optical back-end window of the prototype, is ~ 8.4 mW.

The voltage signal from the cell microphone is amplified by a frequency-selective low-noise amplifier (selective nanovoltmeter type 273, Unipan) and stored by a HS3 digital oscilloscope connected to a personal computer. The microphone signal is stored by the oscilloscope as a time-sample signal realization over a fixed time interval $\tau_l \approx 1.31$ s. The reference signal from the oscilloscope is applied in order to synchronize the triggering of stored realizations. The photoacoustic response is determined as a Fourier transform $S(f_m)$ of the microphone signal $S_t$ for a frequency $f = f_m$ [48]. The quantity $S(f_m)$ is a complex-valued number calculated with the help of a fast-Fourier-transform procedure performed for an individual time-sample signal realization stored in the oscilloscope. In the study we analyze an average of the measured response $S(f_m)$ over $n$ signal-sample realizations. The total time of signal averaging $\tau_{avr}$ for such an averaged quantity is equal to $n\tau_l$.

Here, we analyze photoacoustic signals, which are arisen due to absorption of the laser beam by ammonia. The absorption spectrum for ammonia near a wavelength of 1.53 μm is a group of lines, which belong to overtone and combination band transitions [49]. The lines are overlapped at atmospheric pressures. All our measurements are performed at a fixed laser wavelength $\lambda = 1531.67$ nm, which corresponds to a line-group peak to be observed in the absorption spectrum of ammonia. The laser diode is tuned accurately on this wavelength with the help of the temperature controller. According to [49,50], the ammonia absorption coefficient $\alpha^{(NH3)}$ for this laser wavelength at atmospheric pressure and room temperature (295 K) is 0.34 cm$^{-1}$atm$^{-1}$. We assume that the rates of VV/VT redistribution for the energy to be absorbed by ammonia are high compared to the modulation frequencies realized in the experiment.

All measurements in the experiment are made for the flow of ultra-high purity nitrogen (99.9995 %) produced by a nitrogen generator (ANG250A, Peak Scientific Instruments LTD). The ammonia is produced with the help of a calibrated permeation tube (IM 06-M-A2, Analitpribor, the certified rate of ammonia emission is 1.66 μg/min at 23 degrees Celsius) inserted in a leak-proof box. The rate of nitrogen flow to be blown through the box is maintained automatically near a fixed value with the help of a flow controller. In general, our equipment is capable to generate and control the nitrogen flows at a flow rate in the range from 1.0 up to 500 cm$^3$/min. Therefore, ammonia can be added to the nitrogen flow at a concentration level ranged from ~5 to 5000 parts per million (ppm). In order to reduce the gas flow going through our small-sized prototype cell, a part of the ammonia-contained flow can be taken away through an adjustable relief valve. The rate of flow to be blown through the prototype cell is measured by a glass-tube flow meter with the full scale of 50 cm$^3$/min.

The wall adsorption can play a substantial negative role for detection of ammonia at a ppm-concentration level. Therefore, we make some arrangements which provide reliably the finishing of wall adsorption before the measurements. In order to accelerate the adsorption process, the box PT$_{NH3}$, flow meter FM and cell prototype are connected by short Teflon tubes of small cross section (the length and internal diameter of the tubes are, correspondingly, ~30 cm and 1.2 mm). Before the measurement, the gas line is kept at the required nitrogen-ammonia flow during a long time period sufficient for the absorption to finish. Usually, this period is not shorter than 48 hours. All the measurements show good hour-by-hour and day-by-day reproducibility. The maximal amplitude of photoacoustic response associated with ammonia presence in the gas flow is reproduced within a relative error of 10 %.

The pressure and temperature of gas in the prototype cell are typical for the laboratory room (740 torr and 23 degrees Celsius). Notice that, in the experiments, no capsule is applied for acoustic isolation of the cell. The inlet and outlet gas ducts are opened. A standard day-to-day noise level is kept in the laboratory room. In the room there are no specially-designed sources of external disturbances.

# 4 Frequency dependence of signals

## 4.1 Noise-associated measurement error

An obtained frequency spectrum of standard deviation $\sigma_f$ for the noised microphone signal to be generated inside the acoustically-isolated cell prototype in the absence of light beam is represented in Fig. 4. The spectrum is shown for the frequency range from 3 to 6 kHz. The observed deviation $\sigma_f$ is shown in the figure to be a slowly varying function of $f$. Over the presented frequency range, the magnitude of $\sigma_f$ is in the interval of values from 0.055 to 0.085 μV/Hz$^{1/2}$. We associate this measurement-error value with inherent electric noise of the applied microphone and fluctuations of gas pressure inside the cell. At a frequency $f$ = 4.38 kHz, the quantity $\sigma_f$ is equal to a noise level $\sigma_{f2} \approx$ 0.0795 μV/Hz$^{1/2}$.

## 4.2 Amplitude of laser-initiated signals

We analyze the photoacoustic response to laser-beam absorption inside the prototype cell as a function of modulation frequency $f_m$ for the frequency range from 3 to 6 kHz. The laser-initiated response is analyzed as a sum of components. The components are a parasite background signal (we associated this background, mainly, with absorption of laser beam in the cell windows) and a useful signal (a response to light absorption by gas inside the cell). In order to determine these signal components we perform the measurements for two distinct gas flows (a flow of pure nitrogen or a flow of ammonia-containing nitrogen) blown through the cell. In the measurements, the relief valve RV is closed and the gas flow is blown sequentially through the leak-proof box PT$_{NH3}$, flow meter FM and prototype cell at a rate of 8 cm$^3$/min. This rate value is maintained by a flow-rate controller FC with the full scale of 10 cm$^3$/min. The ammonia concentration in the ammonia-containing flow is $C_I^{(NH3)} \approx 276$ ppm.

Obviously, the response $S(f_m)^{(N2)}$ to be detected for the pure-nitrogen flow is a background signal. In order to reduce the negative influence of the noise on the measurements and to provide a reliable determination of this signal, we perform the averaging of the signal $S(f_m)^{(N2)}$ over an ensemble of multiple ($n \gg 1$) realizations. The response $S(f_m)^{(NH3)}$ observed for the ammonia-containing nitrogen flow is a sum manifestation of the background and useful signals. This response is found as a quantity averaged over $n = 1$ signal-sample realization. The measured amplitudes of the responses $S(f_m)^{(N2)}$ (the number of signal-sample realizations $n$ for this signal is 16) and $S(f_m)^{(NH3)}$ are represented in Fig. 4 as functions of modulation frequency $f_m$.

According to the obtained data, the background response $S(f_m)^{(N2)}$ is a slightly varying function of $f$. Over the presented frequency range, the amplitude $|S(f_m)^{(N2)}|$ is in the interval of values from 0.06 to 0.14 μV. The amplitude-frequency dependence of the background response $S(f_m)^{(N2)}$ exhibits no resonance peak near the eigen-frequency of the second longitudinal acoustic mode.

In the presence of ammonia in the gas flow, the amplitude-frequency dependence of response is essentially transformed. Adding ammonia to the flow leads to an increase in the response amplitude. The observed response amplitude $|S(f_m)^{(NH3)}|$ demonstrates clearly a resonance peak near the eigen-frequency of the $\nu_2$ mode. The peak amplitude $|S(f_2)^{(NH3)}| \approx 43.3$ μV is realized when the modulation frequency is close to a frequency $f_2 \approx 4.38$ kHz. Obviously, the observed peak is a manifestation of the resonance between the modulated laser beam and acoustic $v_2$ mode. For the resonance peak, the Q-factor $Q_2$ ($Q_2 = f_2/\Delta f_2$, the width $\Delta f_2$ is measured between the points where the amplitude is a $1/2^{1/2}$ value of the peak amplitude) is approximately 13.9. The resonance answers to the maximum portion of the useful signal in relation to the noise and background. Definitely, for an ammonia concentration $C_1{}^{(NH3)}$ (a few hundreds of ppm) to be applied in the measurements, the observed response $S(f_m)^{(NH3)}$ at $f_m = f_2$ can be accepted as a useful signal. The ultimate sensitivity of gas detection may be realized for the cell prototype when the modulation frequency $f_m$ is close to the resonance-peak frequency $f_2$.

# 5 Cell performance at acoustic resonance

In order to estimate the performance of the prototype operated in resonance with acoustic $v_2$ mode we analyze properties of laser-initiated photoacoustic signals to be generated at the modulation frequency $f_m = f_2$. Accurate measures of the signal properties are the amplitude and measurement error for the responses $S(f_2)^{(N2)}$ and $S(f_2)^{(NH3)}$ averaged over a sufficiently long time scale $\tau_{avr}$.

## 5.1 Amplitude and measurement error of background signal

The measurement error and amplitude of background signal are evaluated in experiments with a flow of pure nitrogen. In the experiments, the relief valve RV is closed and the nitrogen flow is blown sequentially through the flow meter FM and prototype cell at a rate of 8 cm$^3$/min maintained automatically by a flow-rate controller FC with the full scale of 10 cm$^3$/min. All measurements are done at a laser beam power of 8.36 mW. The time of signal averaging is $\tau_{avr} = 412\tau_1 \approx 540$ s.

The measured amplitude $|S(f_2)^{(N2)}|$ of background response is $\sim 0.13$ µV. We accept this amplitude value as a level of background signal for the cell. The value is represented by a red full square in Fig. 4. According to the obtained data, the fluctuations in the background signal are due to the microphone noise. The observed bandwidth-normalized standard deviation of background signal $\sigma(f_2)^{(N2)} \approx 0.080$ µV/Hz$^{1/2}$ is approximately equal to a minimal microphone-noise level $\sigma_{f2}$, which can be attained (see subsection 3.2) for a frequency $f = f_2$ in the absence of laser beam at negligibly small external acoustic disturbances. The confidence interval $[|S(f_2)^{(N2)}| - \sigma_{f2}, |S(f_2)^{(N2)}| + \sigma_{f2}]$ stated at the 68.2 % confidence level for the background signal to be averaged over $\tau_{avr} = 1$ s is shown as a red error bar of $|S(f_2)^{(N2)}|$ in

Fig. 4. We assume that the minimal microphone-noise level $\sigma_{f2} \approx \sigma(f_2)^{(N2)}$ can be accepted as a noise level of our cell.

## 5.2 Performance of absorption detection

The gas-detection performance of photoacoustic cell is limited usually by noise and background effects. Therefore, we specify the capability of our cell to detect absorption of light beam by gas in terms of absorption coefficients and ammonia concentrations, which are equivalent to the noise or background level obtained for the cell. These equivalents are estimated by evaluating the useful response to light absorption by ammonia-containing gas inside the cell. The estimation is made by two ways. The both ways of performance estimation imply that the ammonia concentrations applied in the estimations are sufficiently high and the observed response $S(f_2)^{(NH3)}$ is approximate to the useful signal. In the first way, the amplitude of useful signal is determined as a response amplitude $|S(f_2)^{(NH3)}|$ evaluated for a gas flow containing ammonia at a fixed concentration. The equivalents are found as quantities proportional to the ratio of noise or background level to the measured amplitude $|S(f_2)^{(NH3)}|$. In the second way, the useful signal is analyzed as a function of ammonia concentration $C^{(NH3)}$. This function is approximated by a dependence determined in measurements of amplitude $|S(f_2)^{(NH3)}|$ at different values of $C^{(NH3)}$. The cell performance is estimated by extrapolating the approximating dependence to ammonia concentrations, at which the amplitude of useful signal is comparable with the noise or background levels.

### 5.2.1. Performance estimation in measurements at a fixed ammonia concentration

The absorption-detection performance of our cell is estimated with the help of measurements for a flow of nitrogen, which contains ammonia at a fixed concentration. In the measurements, the ammonia concentration in the flow is constant and equal to $C_1^{(NH3)} \approx 276$ ppm. The relief valve RV is closed and the gas flow is blown sequentially through the leak-proof box ($PT_{NH3}$), flow meter FM and prototype cell at a rate of 8 cm³/min maintained automatically by a flow-rate controller FC with the full scale of 10 cm³/min. All measurements are done at a laser beam power of 8.36 mW. The time of signal averaging is $\tau_{avr} = 412\,\tau_1 \approx 540$ s.

In the experiment the amplitude and measurement error of useful signal are evaluated. According to our measurements, the amplitude of useful signal $|S(f_2)^{(NH3)}|$ is ~ 43.0 μV. The measured bandwidth-normalized standard deviation for the useful signal is ~ 0.098 μV/Hz$^{1/2}$. The obtained amplitude $|S(f_2)^{(NH3)}|$ is applied then to estimate the absorption-detection performance of cell.

The noise-limited sensitivity of photoacoustic cell to detect absorption is specified usually in terms of a noise-equivalent absorption *NNEA* normalized to laser-beam power and detection bandwidth. This quantity corresponds to a minimal detectable absorption coefficient, at which the signal-to-noise ratio is equal to 1 if the laser-beam power is 1 W and the signal-averaging time is 1 s. At given values of the power $P_{on}$ and signal-to-noise ratio *SNR* realized in the experiment, the quantity *NNEA* can be found from:

$$NNEA = \alpha^{(NH3)} C_l^{(NH3)} P_{on} \, SNR^{-1}. \qquad (2)$$

The signal-to-noise ratio SNR is defined as a ratio of the amplitude of useful response to the bandwidth-normalized noise level $\sigma_{f2}$. According to our measurements at $P_{on}$ = 8.36 mW, the quantity *SNR* is equal to $|S(f_2)^{(NH3)}|/\sigma_{f2} \approx 545$. Correspondingly, the noise-equivalent absorption *NNEA* for the prototype cell is estimated to be ~ $1.44 \times 10^{-9}$ cm$^{-1}$ W Hz$^{-1/2}$.

The noise-equivalent ammonia concentration $C_{NE}^{(NH3)}$, detectable at a typical power $P_{on}^{(typ)}$ of modulated laser beam and a signal-averaging time of 1 s, is

$$C_{NE}^{(NH3)} = NNEA/(\alpha^{(NH3)} P_{on}^{(typ)} \tau_{avr}^{1/2}). \qquad (3)$$

At a power of $P_{on}^{(typ)}$ = 10 mW (a typical power for the near-infrared single-mode laser diodes) the quantity $C_{NE}^{(NH3)}$ is equal to ~ 0.42 ppm.

In comparison to the amplitude of useful response $|S(f_2)^{(NH3)}|$, the background amplitude $|S(f_2)^{(N2)}|$ is considerably low (by a factor of $G \approx 333$). The measured amplitude of the background signal is equivalent, approximately, to an absorption coefficient

$$\alpha_{bg} = \alpha^{(NH3)} C_l^{(NH3)}/G \approx 2.82 \times 10^{-7} \text{ cm}^{-1}. \qquad (4)$$

The background-equivalent concentration of ammonia for the cell prototype is:

$$C_{bg}^{(NH3)} = \alpha_{bg}/\alpha^{(NH3)} \approx 0.83 \text{ ppm} \qquad (5)$$

## 5.2.2. Performance estimation in measurements at multiple ammonia concentrations

The absorption-detection performance of our cell is estimated also with the help of measurements done for a flow of nitrogen containing ammonia at multiple values of concentration $C^{(NH3)}$. In the measurements, the nitrogen flow is blown through the box PT$_{NH3}$ with the permeation tube at a fixed rate ranged from 8 up to 500 cm$^3$/min. This rate value is maintained by a flow-rate controller FC with the full scale of 500 cm$^3$/min. Depending on the maintained flow rate, the ammonia concentration in the flow is a quantity ranged from ~ 5 to 280 ppm. The relief valve RV is opened and adjusted in such a way as to keep the rate of flow through the prototype cell near a constant value (~ 30 cm$^3$/min) regardless of the rate of flow through the box. The measurements are done at a laser-beam power of 8.36 mW.

First of all, we measure the amplitude of $S(f_2)^{(NH3)}$ response at different values of ammonia concentration $C^{(NH3)}$. The amplitude $|S(f_2)^{(NH3)}|$ measured at seven different values of $C^{(NH3)}$ is shown in Fig. 5. The represented values of $|S(f_2)^{(NH3)}|$ are obtained at signal-averaging times $\tau_{avr}$ longer than 30 s. The standard deviation for each of the obtained amplitude values is approximately 0.086 μV/Hz$^{1/2}$.

Then, we evaluate the dependence of useful signal on the quantity $C^{(NH3)}$. In general, the amplitude of useful signal is a linear function of $C^{(NH3)}$. We take into account that, for the response $S(f_2)^{(NH3)}$ measured at ammonia concentrations higher than 5 ppm, the useful-signal component is much stronger in comparison to the background. Therefore, we suppose that the amplitude of useful signal can be approximated by a product $\chi C^{(NH3)}$ where the factor $\chi$ is:

$$\chi = <|S(f_2)^{(NH3)}|/C^{(NH3)}> \approx 0.191 \text{ μV/ppm.}$$

Here the symbol <…> means the averaging over seven experimental points represented in Fig. 5. This figure demonstrates that the product $\chi C^{(NH3)}$ is a good linear approximation function for the measured concentration dependence of $|S(f_2)^{(NH3)}|$.

The noise- and background-equivalent ammonia concentrations can be defined as concentrations, at which the product $\chi C^{(NH3)}$ is equal, correspondingly, to the standard deviation $\sigma_{f2}$ and amplitude $|S(f_2)^{(N2)}|$ (quantities specifying the noise and background levels). Graphically, such concentrations are the abscissas for points of intersection between the approximation line $\chi C^{(NH3)}$ with horizontal lines, the magnitude of which are equal to $\sigma_{f2}$ or $|S(f_2)^{(N2)}|$. In accordance with the data represented in Fig. 5, the noise- and background-equivalent ammonia concentrations equal, correspondingly, $\sigma_{f2}/\chi \approx 0.42$ ppm and $|S(f_2)^{(N2)}|/\chi \approx 0.68$ ppm. These values are in good agreement with the relevant quantities $C_{NE}^{(NH3)}$ and $C_{bg}^{(NH3)}$ estimated for the typical power $P_{on}^{(typ)} = 10$ mW in subsection 5.2.1. We can conclude that the measurements done at multiple ammonia concentrations confirm the validity of performance parameters estimated for a fixed concentration in subsection 5.2.1.

## 5.3 Performance of gas-leak detection

We estimate the capability of the produced cell prototype to detect the flow of an absorbing gas-tracer. In the estimation, we assume that the flow is emitted by an individual small-sized object in the environment of a non-absorbing carrier gas. The carrier gas containing the gas-tracer is blown through the central cylindrical part of cell (the volume of this part is a quantity $\pi(D/2)^2 L_C \approx V/2$) at a flow rate $(V/2)/\tau_{renew}$, which answers to a gas renewal inside this cell part during the time of $\tau_{renew}$. We assume that the blown gas-tracer is concentrated, mainly, within the central part of cell cavity. In the estimation, we suppose that the modulated laser beam has a power $P_{on}^{(typ)} = 10$ mW (a typical power of laser beam for the near-infrared single-mode laser diodes). We accept also that the leak detection are performed at a standard time resolution of 1 s and $\tau_{avr} = \tau_{renew} = 1$ s. The estimation implies the usage of performance parameters (they are the quantities $NNEA$, $C_{NE}^{(NH3)}$ and $\alpha_{bg}$ found in subsection 5.2.2) evaluated in measurements for a nitrogen flow containing ammonia at a fixed concentration $C_1^{(NH3)}$.

The noise-limited sensitivity of gas-leak detection can be specified in terms of a noise-equivalent cross-section ($NNECS$) normalized to the laser-beam power and detection bandwidth. For the banana- or Π-shaped photoacoustic cells, this quantity is

$$NNECS = \pi(D/2)^2 L_C NNEA = (V/2)NNEA. \qquad (6)$$

The quantity $NNECS$ specifies a noise-limited sensitivity of gas-leak detection for the photoacoustic cell without reference to the absorption properties of substance to be detected. For the produced cell prototype, the cross-section $NNECS$ equals approximately $1.43 \times 10^{-10}$ cm$^2$ W Hz$^{-1/2}$.

A noise-equivalent rate $R_{NE}^{(NH3)}$ of ammonia flow gives the minimal noise-limited leak rate detectable at fixed values of times $\tau_{avr}$, $\tau_{renew}$ and power $P_{on}^{(typ)}$. For the banana- or Π-shaped cells, this quantity is:

$$R_{NE}^{(NH3)} = \pi(D/2)^2 L_C C_{NE}^{(NH3)}/\tau_{renew} = NNECS/(\alpha^{(NH3)}P_{on}^{(typ)}\tau_{avr}^{1/2})/\tau_{renew}. \qquad (7)$$

According to the obtained data and selected values of $\tau_{avr}$, $\tau_{renew}$ and $P_{on}^{(typ)}$, the produced cell prototype is specified by a rate $R_{NE}^{(NH3)} \approx 4.20 \times 10^{-8}$ cm$^3$/s.

The effect of background signal on the leak-detection performance can be evaluated in terms of a volumetric amount of detected gas, which enters into the cell and produces a photoacoustic response equivalent to the background signal. For ammonia blown through the central cavity part of our cell, this quantity is equal to

$$V_{bg}{}^{(NH3)} = \pi(D/2)^2 L_C \alpha_{bg}/\alpha^{(NH3)} = S_{bg}/\alpha^{(NH3)}. \qquad (8)$$

Here the quantity

$$S_{bg} = \pi(D/2)^2 L_C \; \alpha_{bg} \approx (V/2)\alpha_{bg} \qquad (9)$$

is defined as a background-equivalent cross-section. Similar to *NNECS*, the quantity $S_{bg}$ specifies the performance of gas-leak detection for the photoacoustic cells irrespective of the substance to be detected. For the produced cell prototype, the quantities $S_{bg}$ and $V_{bg}{}^{(NH3)}$ are correspondingly $\sim 2.80 \times 10^{-8}$ cm$^2$ and $\sim 8.23 \times 10^{-8}$ cm$^3$.

# 6. Discussion

The performance parameters *NNEA*, $C_{NE}{}^{(NH3)}$, $\alpha_{bg}$, $C_{bg}{}^{(NH3)}$, *NNECS*, $R_{NE}{}^{(NH3)}$, $S_{bg}$ and $V_{bg}{}^{(NH3)}$ obtained for the produced prototype cell are represented in Table 1. For reference, this table contains also the relevant performance parameters estimated with the help of Eqs. (2)-(9) for the banana-shaped cells [40, 41].

## 6.1. Performance of absorption detection

In comparison to the miniature prototype [41] of banana-shaped cell, the produced prototype of Π-shaped cell demonstrates a considerably high performance to detect absorption in gas. The normalized noise-equivalent absorption *NNEA* is lower for the produced cell prototype than for the cell [41] by a factor of $A \approx 56$. The low value of noise-equivalent absorption is explained by several reasons. First of all, the acoustic sensor mounted in the prototype is a condenser microphone, the signal-to-noise ratio of which is $\sim 3$ times higher than one of a MEMS-based ultrasonic transducer applied in the cell [41]. For adapting the prototype cell to operation with this low-frequency acoustic sensor, the acoustic cavity of cell is essentially (by a factor of $\sim 8$ in comparison with the cavity [41]) lengthened. Evidently, contrary to the transducer of the cell [41], the microphone of our prototype is well acoustically coupled with the cavity.

At the resonance between the modulated laser beam and acoustic $v_2$ mode, the absorption equivalent of background signal for the produced prototype is ~ 88.7 times lower than one for the cell [41]. Over the frequency range from 3 to 6 kHz, the background amplitude does not depend on the modulation frequency $f_m$. We associate the low value of background amplitude for our cell with a slight effect of imperfections in the cavity design, cell-manufacturing process and cell-alignment procedure. The cell design is optimized with the help of an accurate numeric simulation. The cell production is facilitated by a simpler design and larger sizes of acoustic cavity. The proper alignment of cell along the laser beam is simplified by a larger cross-section diameter $D$ of cavity.

Despite an evident improvement in absorption-detection sensitivity for the miniaturized cells, the produced prototype cell is still specified by a low performance of absorption sensing in comparison with the macro-scale banana-shaped cell [40]. The noise-equivalent absorption $NNEA$ is higher for the produced prototype than for the cell [40] by a factor of ~ 7.2. The background-equivalent absorption $\alpha_{bg}$ for the prototype is ~ 14.1 times higher than one for the cell [40].

## 6.2. Performance of gas-leak detection

In comparison with the referenced banana-shaped cells [40,41], the produced prototype cell demonstrates a considerably better performance of gas-leak detection. In spite of larger cell sizes (the volume of acoustic cavity for the prototype is ~ 40 times larger than for the miniature cell [41]), the normalized noise-equivalent cross-section $NNECS$ and ammonia-leak rate $R_{NE}^{(NH3)}$ are lower for the prototype than for the cell [41] by a factor of ~ 1.4. The background-equivalent cross-section $S_{bg}$ and volumetric amount of ammonia $V_{bg}^{(NH3)}$ for our cell are ~ 2.3 times lower than ones for the cell [41]. The produced cell prototype shows a substantially better performance of gas-leak detection in comparison both to the macro-scale banana-shaped cell and to the commercial portable gas-leak detectors. For comparison, the minimal detectable leak rate for commercial hand-held halogen/hydrogen/helium sniffer leak detectors attains a value down to ~ $10^{-6}$ cm$^3$/s [18-22].

# 7 Conclusion

Thus, we have presented a simple design of acoustic cavity for the photoacoustic cell. The cavity shape is accurately optimized so to minimize the window background and enhance the gas-detection performance for the cell operated at a selected acoustic resonance. The presented cavity shape is implemented in a compact prototype cell designed and produced in accordance with recommendations elaborated previously in [41]. The detailed experimental examination demonstrates a good capability for the produced cell both to measure the trace concentrations of chemical compounds in gas media and to sense the weak gas leak emitted by small-scale individual objects.

The obtained results demonstrate a great potential for the miniaturized traditionally-designed resonant photoacoustic cells in creating hand-held high-sensitivity laser-spectroscopy sensors of chemical compounds. The prototype cell equipped with a miniature near-infrared laser can be used as a low-cost basis in order to develop a 'pocket-sized' gas detector capable of sensing the smallest gas leaks. In combination with amplitude- or wavelength-modulated laser beams, such a cell is capable of high-sensitivity detection of any infrared-active chemical compound whatever the spectral features of the compound. The potential of this cell in detection of complex polyatomic compounds with broad spectral absorption bands will be demonstrated in the nearest future.

# List of table and figure captions

Table 1 Design, resonance properties and gas-detection performance for banana-shaped cells [40,41] and our cell prototype.

Fig. 1 Design of acoustic cavity of photoacoustic cell: ($OO'$) optical axis; ($L_c$) length of the central cylindrical part; ($L_{lat}$) length of the lateral cylindrical parts; ($D$) cross-section diameter for the central and lateral cylindrical parts; ($\Theta_B$) angle between $OO'$ and the normal to the window surface; ($M$) acoustic sensor. Location of the inlet/outlet gas holes is shown by stars.

Fig. 2 Photo of the produced cell prototype: (1, 1') optical front- and back-end windows; (2) nipples of inlet/outlet gas holes.

Fig. 3 Experimental setup for analysis of laser-initiated photoacoustic signals to be generated inside the produced cell prototype: ($N_2$-gen) nitrogen generator; (FC) gas-flow controller; ($PT_{NH3}$) box with permeation tube; (FM) flow meter; (RV) relief valve; (LDC) current controller; (TED) temperature controller; (DFB) fiber-pigtailed laser diode; (C) fiber collimator;

(λ/2) half-wave plate; (PAC) photoacoustic cell; (PM) power meter; (Amp) frequency-selective amplifier; (DO) digital oscilloscope; (PC) personal computer.

Fig. 4 Noise-associated measurement error and amplitudes of laser-initiated signals as functions of modulation frequency $f_m$ for the produced cell prototype. The black solid line shows the frequency spectrum of deviation $\sigma_f$ for the noised microphone signal to be generated inside the acoustically-isolated cell in the absence of light beam. The magenta open circles show the frequency dependence of background-signal amplitude $|S(f_m)^{(N2)}|$ observed for a flow of pure nitrogen at $\tau_{avr} \approx 21$ s and $P_{on} \sim 8.4$ mW. The blue dashed line demonstrates the frequency dependence of signal amplitude $|S(f_m)^{(NH3)}|$ obtained at a nitrogen flow containing 276 ppm of ammonia ($\tau_{avr} \approx 1.31$ s, $P_{on} \sim 8.4$ mW). The red full square gives the amplitude of background response at $f_m = f_2$ ($\tau_{avr} \approx 540$ s, $P_{on} \approx 8.36$ mW). The red error bar shows the confidence interval $[|S(f_2)^{(N2)}| - \sigma_{f2}, |S(f_2)^{(N2)}| + \sigma_{f2}]$ for the background signal averaged at $\tau_{avr} = 1$ s. The vertical green dash-dot line answers to location of resonance frequency $f_2$ on the abscissa axis.

Fig. 5 Dependence of signal amplitude on the ammonia concentration $C^{(NH3)}$. The values of amplitude $|S(f_2)^{(NH3)}|$ measured at seven different concentrations $C^{(NH3)}$ ($\tau_{avr} > 30$ s, $P_{on} \approx 8.36$ mW) are shown by blue open circles. The linear approximation of concentration dependence for the amplitude of useful signal is given by a blue solid line. The background level specified by the amplitude $|S(f_2)^{(N2)}|$ is represented by a horizontal red dash-dot line. The noise level specified by the deviation $\sigma_{f2}$ is shown as a horizontal fat black dashed line.

| Parameter | Macro-scale banana-shaped cell [40] | Miniature banana-shaped cell [41] | Our Π-shaped cell |
|---|---|---|---|
| **Design** | | | |
| Absorption-path length $L_c$, mm | 300 | 5 | 39 |
| Cross-section diameter $D$, mm | 15 | 0.8 | 1.8 |
| Volume of acoustic cavity, $cm^3$ | 108 | 0.005 | 0.2 |
| Weight(material), g | – | 9.8(brass) | 3.5(plastic) |
| **Resonance properties** | | | |
| Resonance frequency, kHz | 0.55 | 32.9 | 4.38 |
| Q-factor | 55 | 16.3 | 13.9 |
| **Performance of absorption detection ($P_{on}^{(typ)}$ = 10 mW, $\tau_{avr}$ = 1 s, $\alpha^{(NH3)}$ = 0.34 $cm^{-1}atm^{-1}$)** | | | |
| Normalized noise-equivalent absorption $NNEA$, $cm^{-1}$ W $Hz^{-1/2}$ | $2\times10^{-10}$ | $8.1\times10^{-8}$ | $1.44\times10^{-9}$ |
| Noise-equivalent ammonia concentration $C_{NE}^{(NH3)}$, ppm | 0.058 | 23.7 | 0.42 |
| Background-equivalent absorption $\alpha_{bg}$, $cm^{-1}$ | $2\times10^{-8}$ | $2.5\times10^{-5}$ | $2.82\times10^{-7}$ |
| Background-equivalent ammonia concentration $C_{bg}^{(NH3)}$, ppm | 0.058 | 73.8 | 0.83 |
| **Performance of gas-leak detection ($P_{on}^{(typ)}$ = 10 mW, $\tau_{avr}$ = 1 s, $\alpha^{(NH3)}$ = 0.34 $cm^{-1}atm^{-1}$)** | | | |
| Normalized noise-equivalent cross-section $NNECS$, $cm^2$W $Hz^{-1/2}$ | $1.06\times10^{-8}$ | $2.03\times10^{-10}$ | $1.43\times10^{-10}$ |
| Noise-equivalent ammonia-leak rate $R_{NE}^{(NH3)}$, $cm^3/s$ | $3.12\times10^{-6}$ | $5.96\times10^{-8}$ | $4.20\times10^{-8}$ |
| Background-equivalent cross-section $S_{bg}$, $cm^2$ | $1.06\times10^{-6}$ | $6.31\times10^{-8}$ | $2.80\times10^{-8}$ |
| Background-equivalent volumetric amount of ammonia $V_{bg}^{(NH3)}$, $cm^3$ | $3.12\times10^{-6}$ | $1.86\times10^{-7}$ | $8.23\times10^{-8}$ |

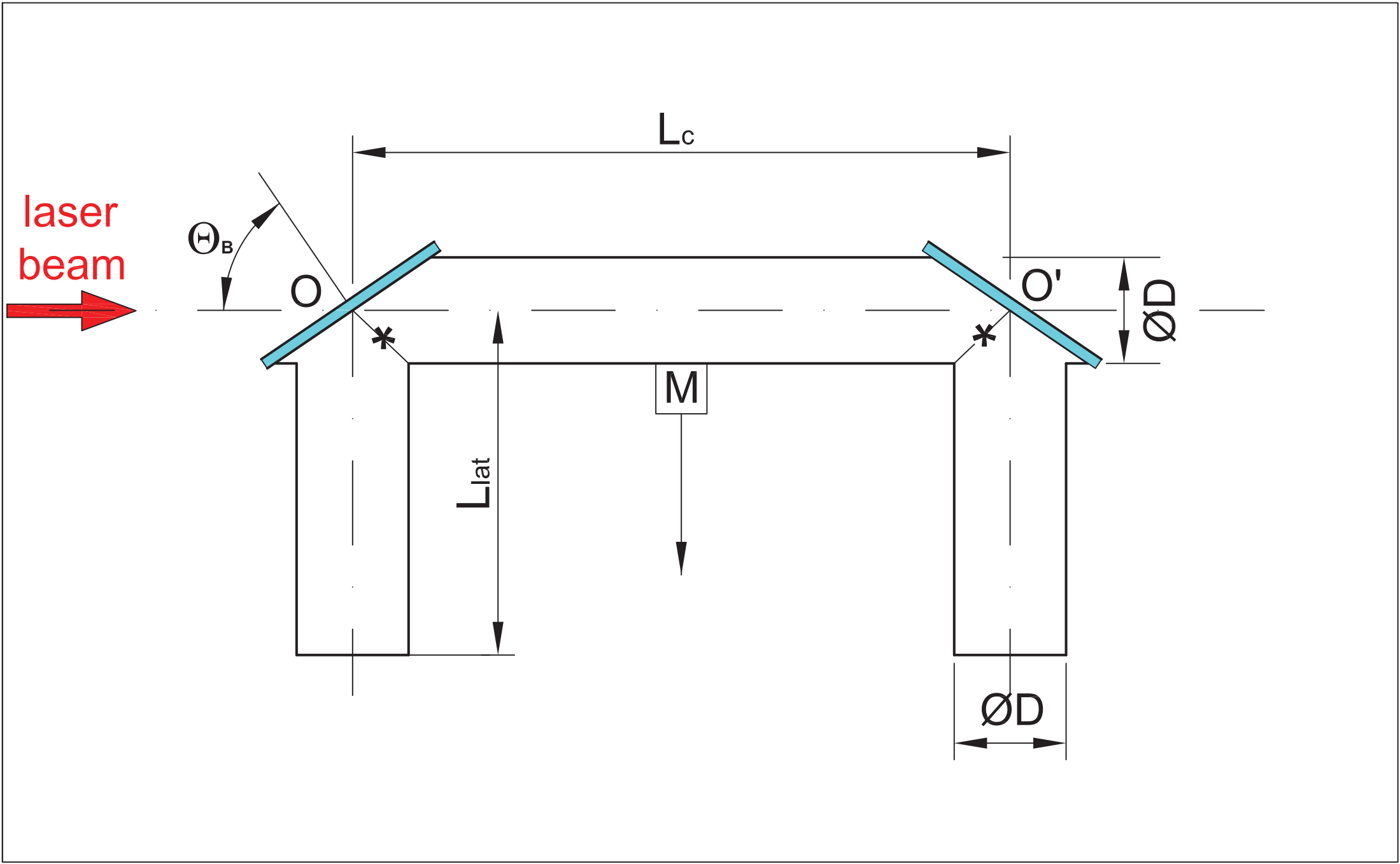

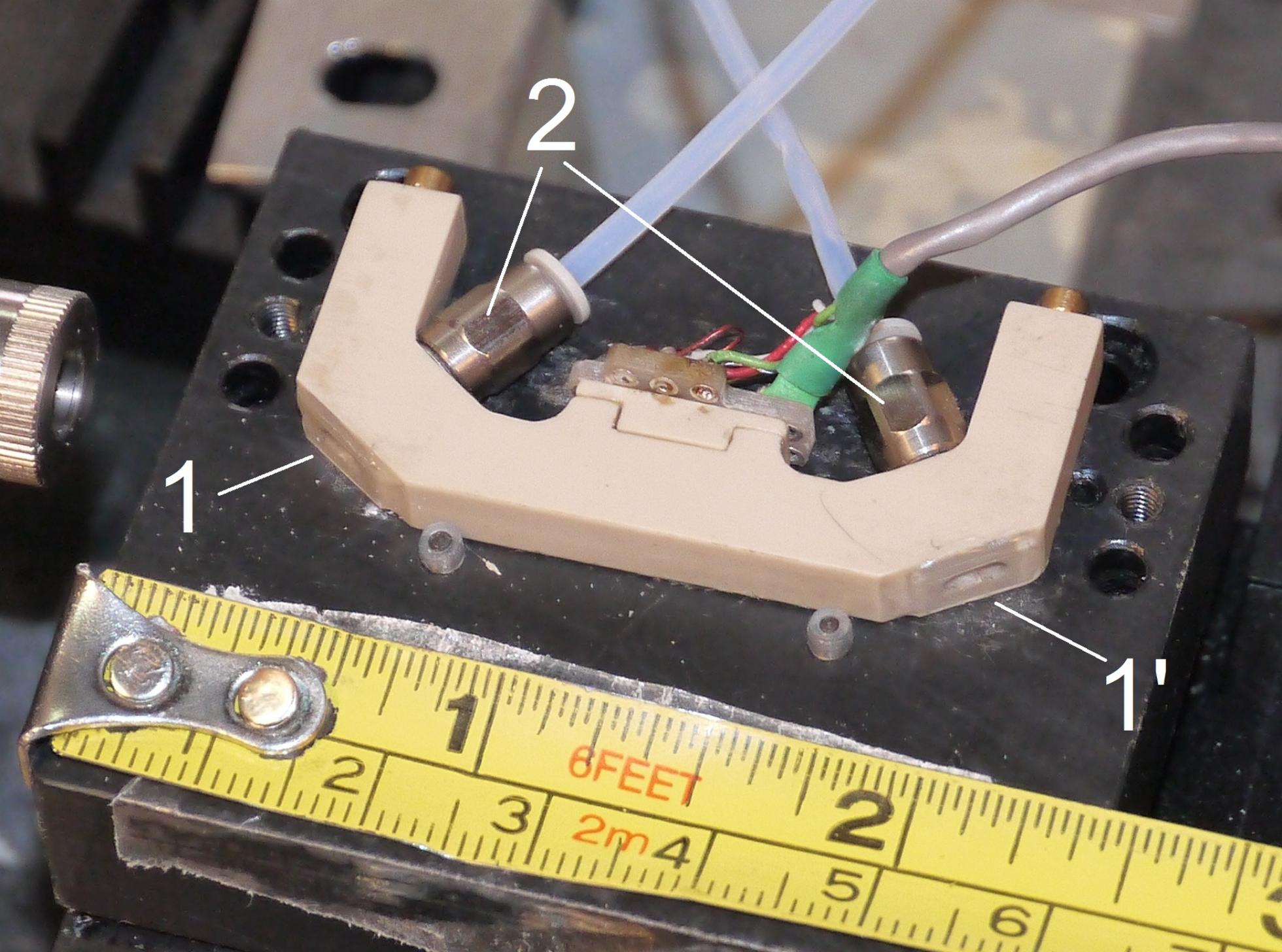

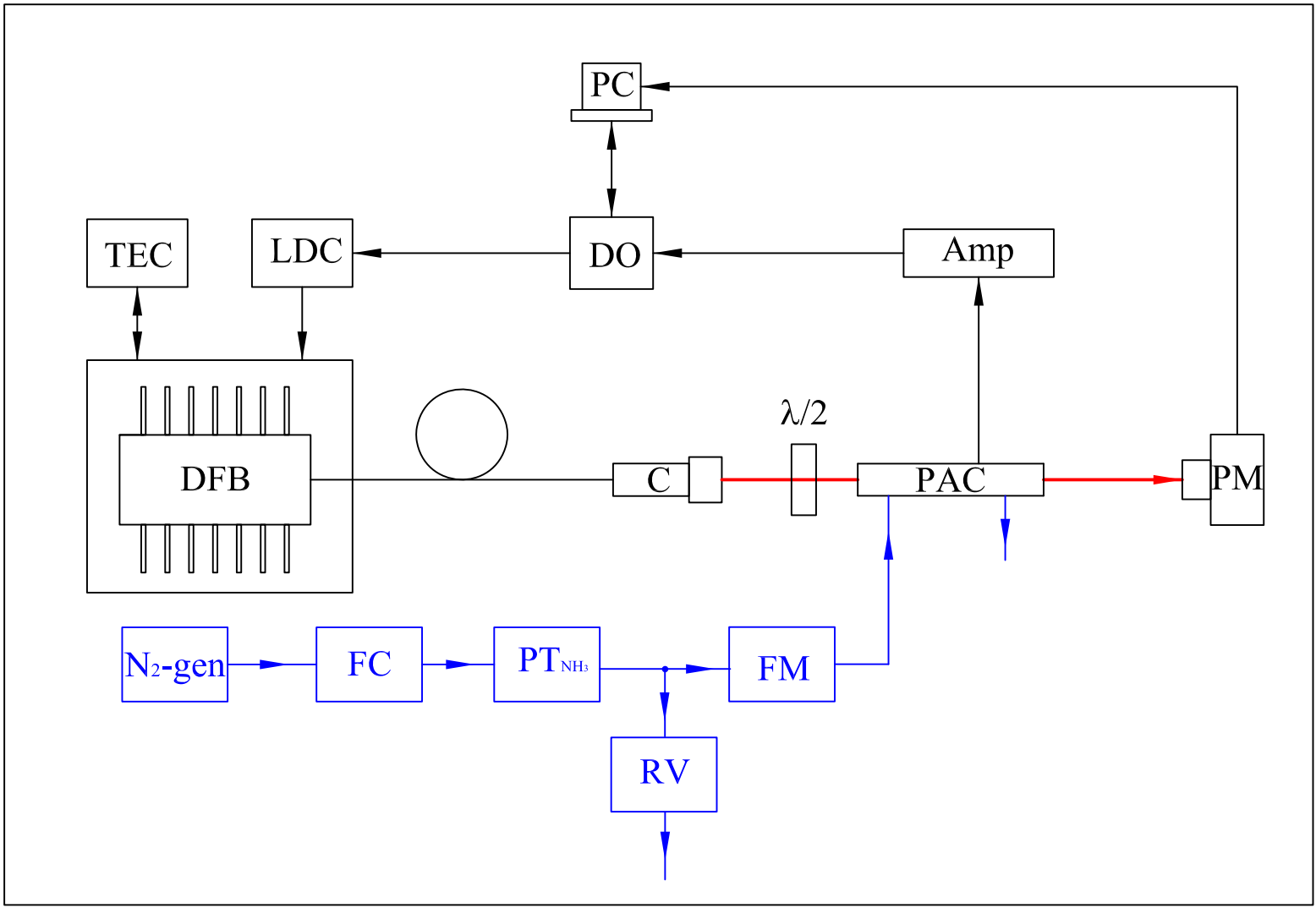

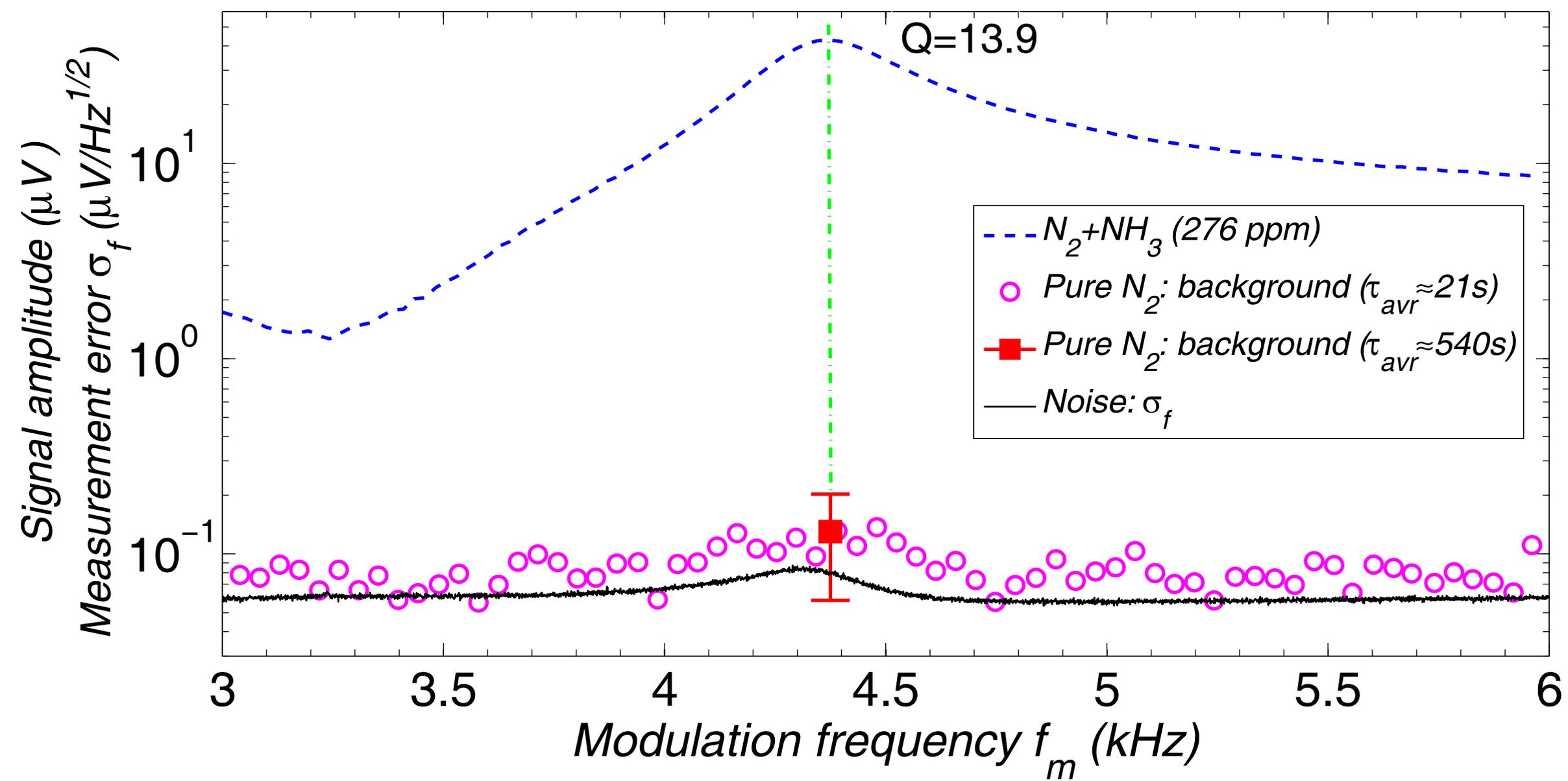

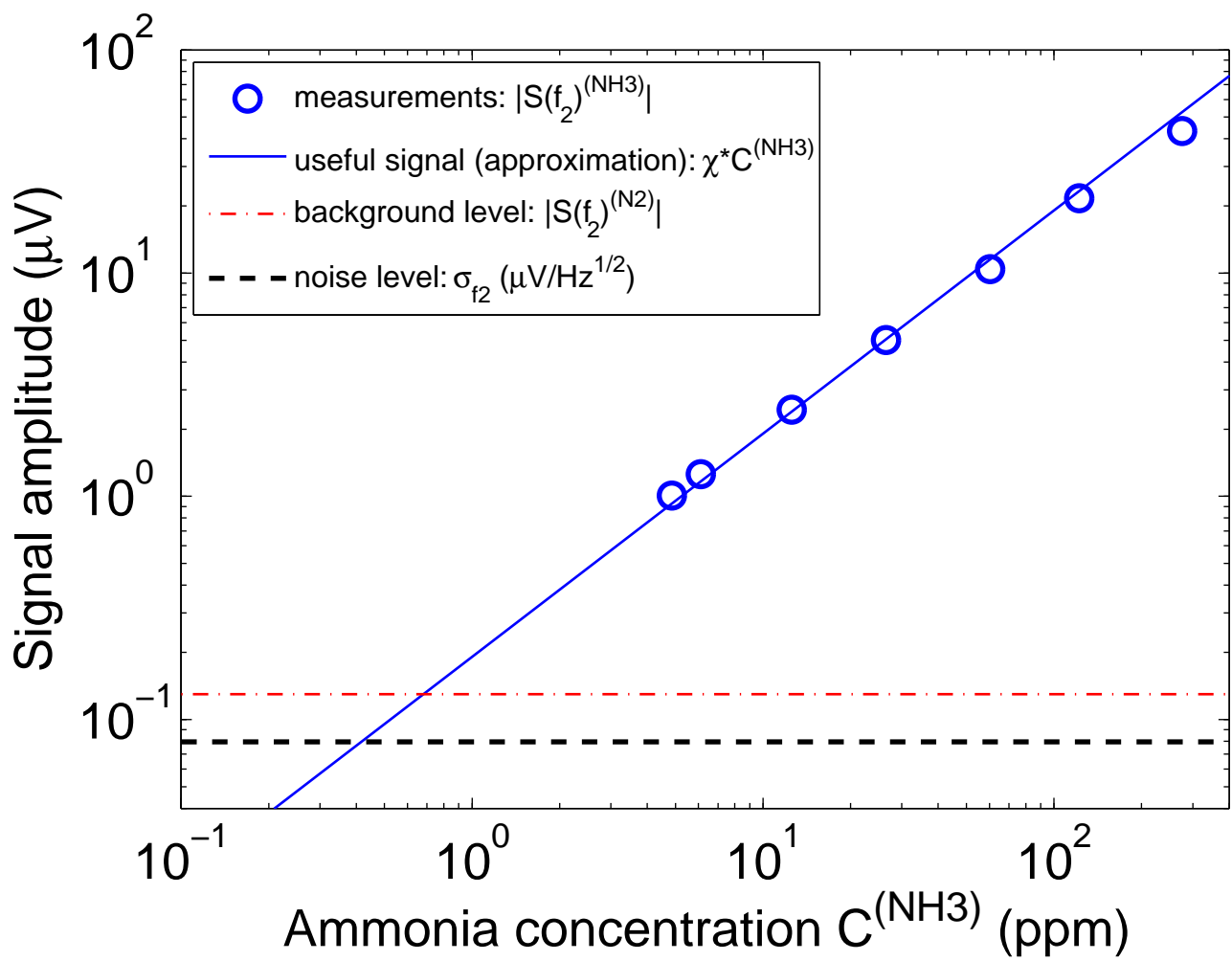